\documentclass[aps,prl,twocolumn,superscriptaddress]{revtex4-1}
\usepackage{amssymb}
\usepackage{graphicx}
\usepackage{graphics}
\usepackage{amsmath}
\usepackage{placeins}
\usepackage{hyperref}
\usepackage{verbatim}

\begin{document}

\title{\emph{In situ} Observation of Dark Current Emission in a High Gradient RF Photocathode Gun}
\author{\firstname{Jiahang} \surname{Shao}}
\email{shaojh07@mails.tsinghua.edu.cn}
\affiliation{Department of Engineering Physics, Tsinghua University Beijing 100084, P.R.China}
\affiliation{Argonne National Laboratory, Lemont, IL 60439, USA}
\author{\firstname{Sergey P.} \surname{Antipov}}
\affiliation{Argonne National Laboratory, Lemont, IL 60439, USA}
\affiliation{Euclid Techlabs LLC, Bolingbrook, IL 60440, USA}
\author{\firstname{Sergey V.} \surname{Baryshev}}
\affiliation{Argonne National Laboratory, Lemont, IL 60439, USA}
\affiliation{Euclid Techlabs LLC, Bolingbrook, IL 60440, USA}
\author{\firstname{Huaibi} \surname{Chen}}
\affiliation{Department of Engineering Physics, Tsinghua University Beijing 100084, P.R.China}
\author{\firstname{Manoel} \surname{Conde}}
\affiliation{Argonne National Laboratory, Lemont, IL 60439, USA}
\author{\\\firstname{Wei} \surname{Gai}}
\affiliation{Argonne National Laboratory, Lemont, IL 60439, USA}
\author{\firstname{Gwanghui} \surname{Ha}}
\affiliation{Argonne National Laboratory, Lemont, IL 60439, USA}
\author{\firstname{Chunguang} \surname{Jing}}
\affiliation{Argonne National Laboratory, Lemont, IL 60439, USA}
\affiliation{Euclid Techlabs LLC, Bolingbrook, IL 60440, USA}
\author{\firstname{Jiaru} \surname{Shi}}
\email{shij@tsinghua.edu.cn}
\affiliation{Department of Engineering Physics, Tsinghua University Beijing 100084, P.R.China}
\author{\firstname{Faya} \surname{Wang}}
\affiliation{SLAC National Accelerator Laboratory, Menlo Park, CA 94025, USA}
\author{\firstname{Eric} \surname{Wisniewski}}
\affiliation{Argonne National Laboratory, Lemont, IL 60439, USA}

\date{\today}
\begin{abstract}
Undesirable electron field emission (a.k.a. dark current) in high gradient RF photocathode guns deteriorates the quality of photoemission current and limits the operational gradient. To improve the understanding of dark current emission, a high-resolution ($\sim$100 $\mu$m) dark current imaging experiment has been performed in an L-band photocathode gun operating at $\sim$100 MV/m of surface gradient. Dark current from the cathode has been observed to be dominated by several separated strong emitters. The field enhancement factor, $\beta$, of selected regions on the cathode has been measured. The post scanning electron microscopy (SEM) and white light interferometer (WLI) surface examinations reveal the origins of $\sim$75\% strong emitters overlap with the spots where rf breakdown have occurred.
\end{abstract}

\maketitle

Electrons can tunnel through a surface barrier modified by the presence of an electric field, resulting in a field emission (FE) current~\cite{FNformula1928,GadzukRMP1973,JuwenSLAC1997,Fursey2007}. While the existence of this physical phenomenon allows the operation of field emission electron sources~\cite{KennethNature2005,XiangkunPRST2013,PiotAPL2014,BaryshevAPL2014}, it has a negative (parasitic) impact on the performance of vacuum resonator-based dc and rf systems such as traveling wave tubes, photocathode guns, and particle accelerators~\cite{JuwenSLAC1997,HanPRST2005,GrudievPRST2009,DescoeudresPRST2009,FayaPRST2011}. The troublesome field emission current is referred to as dark current. It is an incoherent source of electrons that impacts the energy budget of a device, and is a source of undesired secondary electrons and ions~\cite{HanPRST2005,XiangPRST2014,HuangPRST2015}. Historically, dark current has been considered to be a trigger of breakdown in vacuum devices which may interrupt the normal operation of the device and even jeopardize the entire facility~\cite{JuwenSLAC1997}. 

To date many questions surrounding FE still remain, especially in the rf case which limit the improvement of electron sources and high gradient accelerators for TeV-scale linear colliders~\cite{CLICCDR} and compact X-ray electron sources~\cite{SACLANaturePh2012,SwissFELCDR}. For example, a large discrepancy exists between emitter properties obtained through direct observation using advanced surface analysis tools and those indirectly obtained from fitting the experimental data to the Fowler-Nordheim (F-N) equation~\cite{JuwenSLAC1997,HuaibiPRL2012}; the temporal evolution of the FE area under high electromagnetic fields is mostly unknown~\cite{JiahangIPAC2014}; and empirical methods and procedures to suppress or enhance dark current lack theoretical support. All these questions result from the lack of a means for \emph{in situ} high-resolution FE observation. In earlier FE studies under a dc field, emitter mapping with better than 1 $\mu$m resolution has been achieved by scanning an anode along the cathode~\cite{Noer1986JAP,LysenkovIJN2005,PandeyPRST2009}. However, imaging the field emitters at high resolution while they are emitting under an rf field is extremely challenging due to the wide emitting phase (the timing with respect to the applied rf field) and energy spread range of the dark current~\cite{HanPAC2005,MorettiPRST2005,DowellPAC2007,XiangPRST2014}. In this Letter, we present observations of \emph{in situ} dark current emission in a high gradient photocathode gun using a dedicated dark current imaging beamline.

\begin{figure}[h!tbp]
	\includegraphics[width=8cm]{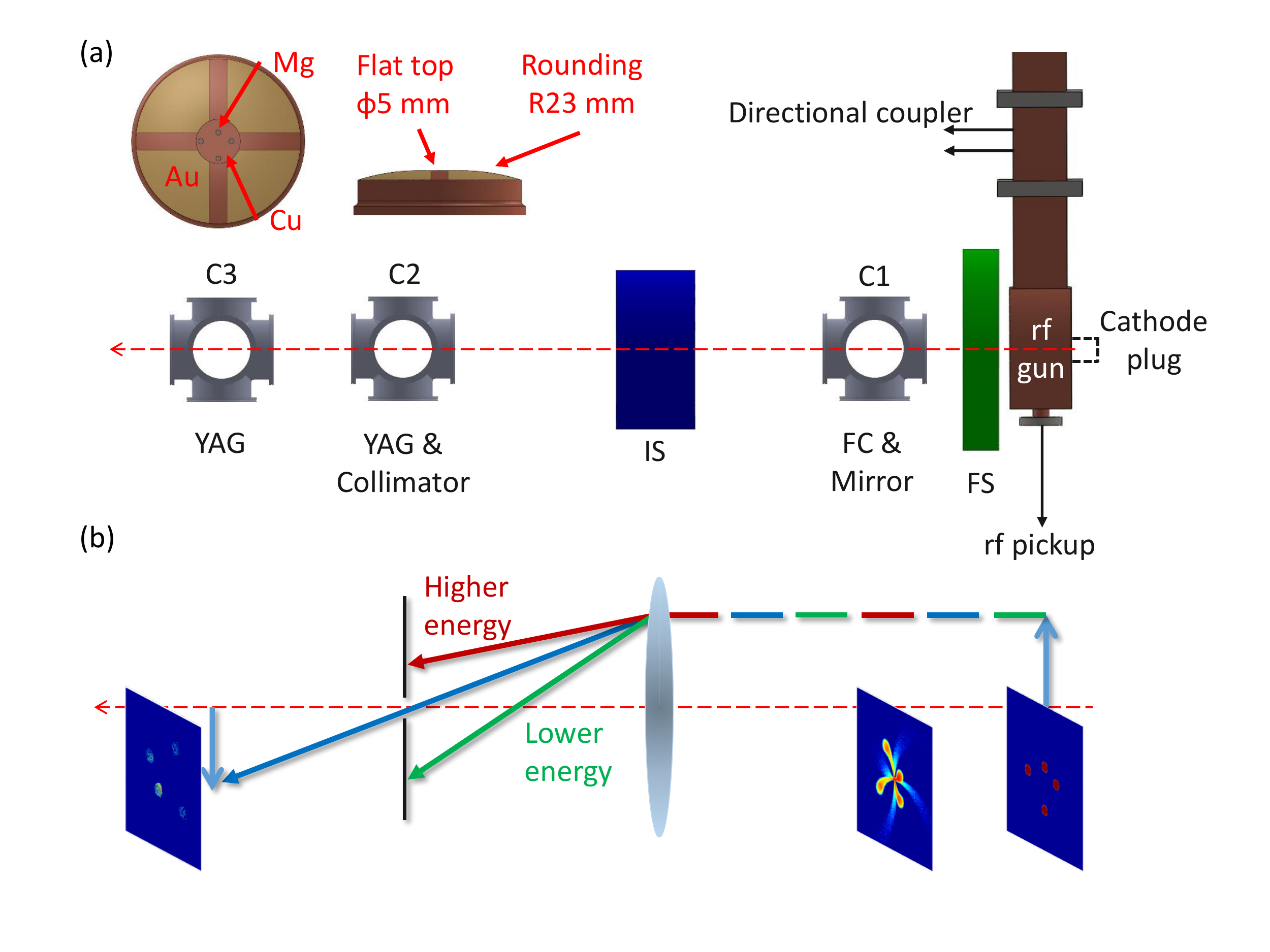}
	\caption{\label{figureSetup}The dark current imaging system at Argonne Wakefield Accelerator facility. (a) The designed cathode with a reference pattern for the field emission images and the beamline layout: FC, Faraday Cup to monitor emission current; C, vacuum Cross to house imaging components which were mounted on retractable actuators; FS, Focusing Solenoid; IS, Imaging Solenoid; and YAG, doped Yttrium Aluminum Garnet phosphor screen. Inset: Top and side view of the novel shaped cathode with its sputtering pattern. (b) The equivalent optical imaging system. Insets: ASTRA simulation results for the emission patterns on the cathode, at the gun exit, and in the imaging plane.}
\end{figure}

The study was conducted at Argonne Wakefield Accelerator facility (AWA). The imaging beamline is shown in Fig. ~\ref{figureSetup}(a)~\cite{JiahangPRL2015}. To achieve high-resolution dark current imaging, a method to select electrons from certain emitting phase and narrow the energy spread was developed using external axial magnetic fields (i.e. solenoids) and a collimator at the focal plane. The object being imaged was a novel-shaped copper cathode in a 1.3 GHz rf gun. The cathode is ~20 cm in diameter with a large edge rounding and a small flat center (inset, Fig. ~\ref{figureSetup}(a)) to enhance FE on the top area. $\sim$100 nm thick magnesium and gold (Mg has a work function of 3.7 eV, Au 5.1 eV, and Cu 4.6 eV) have been sputtered in certain areas to create a spatial pattern as reference. Electrons emitted from the cathode gain energy from the rf gun depending on the emitting phase. They are accordingly focused by the solenoids at different longitudinal positions~\cite{JiahangCLICWS2016}. The transverse positions of electrons depend on their emission phases and applied focusing forces. Thus, a blurred pattern is formed at the exit of the gun and deteriorates downstream, as simulated by the beam dynamics code ASTRA~\cite{ASTRA}. When a collimator with a small aperture is applied after the focusing elements, only electrons with the proper focusing position and energy gain are allowed to pass through. A sharp image can be then obtained. The whole imaging system also can be considered as an optical system, as illustrated in Fig. ~\ref{figureSetup}(b). 

The average magnification and rotation of the imaging system can be defined as
\begin{equation*}
\left\{
\begin{aligned}
\overline{mag}&=\frac{\overline{\rho}}{\rho_{0}}\\
\overline{rot}&=\overline{\varphi}-\varphi _{0}\\
\end{aligned}
\right.
\end{equation*}
where ($\rho_{0}$,$\varphi _{0}$) is the initial emitter position on the cathode in polar coordinates, ($\rho$,$\varphi$) is the image position on the last YAG screen (C3 in Fig. \ref{figureSetup}(a)) of electrons emitted at different phases and transverse angles, and $\overline{\rho}$ and $\overline{\varphi}$ are the average value of $\rho$ and $\varphi$. As the system is axial symmetry, the resolution can be defined in radial and angular direction. Assuming $\rho$ and $\varphi$ follow the Gaussian distribution, the resolutions are defined as
\begin{equation*}
\left\{
\begin{aligned}
R_{\rho}=2.35 \times \dfrac{\delta _{\rho}}{\overline{mag}}\\
R_{\varphi}=2.35 \times \delta _{\varphi}\rho _{0}\\
\end{aligned}
\right.
\end{equation*}
where $\delta _{\rho}$ and $\delta _{\varphi}$ are the standard deviation of $\rho$ and $\varphi$. The resolution improves when smaller apertures are imposed. 

Four 60 $\mu$m thick apertures at a 30 mm spatial interval are mounted on a stainless steel plate which can be precisely moved along the transverse direction by a motorized actuator so as to choose different apertures. The diameters of the apertures are 8 mm, 1 mm, 0.5 mm, and 0.2 mm, respectively. Based on the simulations, 40-140 $\mu$m resolution can be achieved depending on the initial FE electron emittance when the smallest aperture is applied~\cite{JiahangCLICWS2016}.

Diagnostics required in the experiment are a bidirectional coupler to monitor the input and reflected rf signals, an antenna (rf pickup probe) to monitor the rf signal inside the cavity, and a mirror to roughly locate the breakdown position during rf conditioning. The YAG screens are placed perpendicular to the beamline and the image is transported out of the beamline in the transverse direction by a mirror angled at 45$^{\circ}$ and located behind each screen. A PI-MAX Intensified CCD (ICCD) camera~\cite{PI-MAX} is used to capture the image with 47 $\mu$m/pixel resolution on the last YAG screen. Given the image magnification of 5 from the cathode to the YAG screen, the camera resolution at the cathode is 9.4 $\mu$m which does not limit the imaging quality of the whole system.

Before the imaging experiment, the electric field on the cathode (noted as $E_{c}$) has been carefully conditioned to 120 MV/m with $\sim$2.5 MW input power (pulse length was 6.5 $\mu$s). Judged by the flash observed on the mirror~\cite{WalterEPAC2002}, breakdowns occurred on the cathode and inside the cavity. After the conditioning, $E_{c}$ was lowered to 105 MV/m. Steady dark current emission regions on the metal surface were observed and no further breakdowns occurred on those areas.

\begin{figure}[h!tbp]
	\includegraphics[width=8cm]{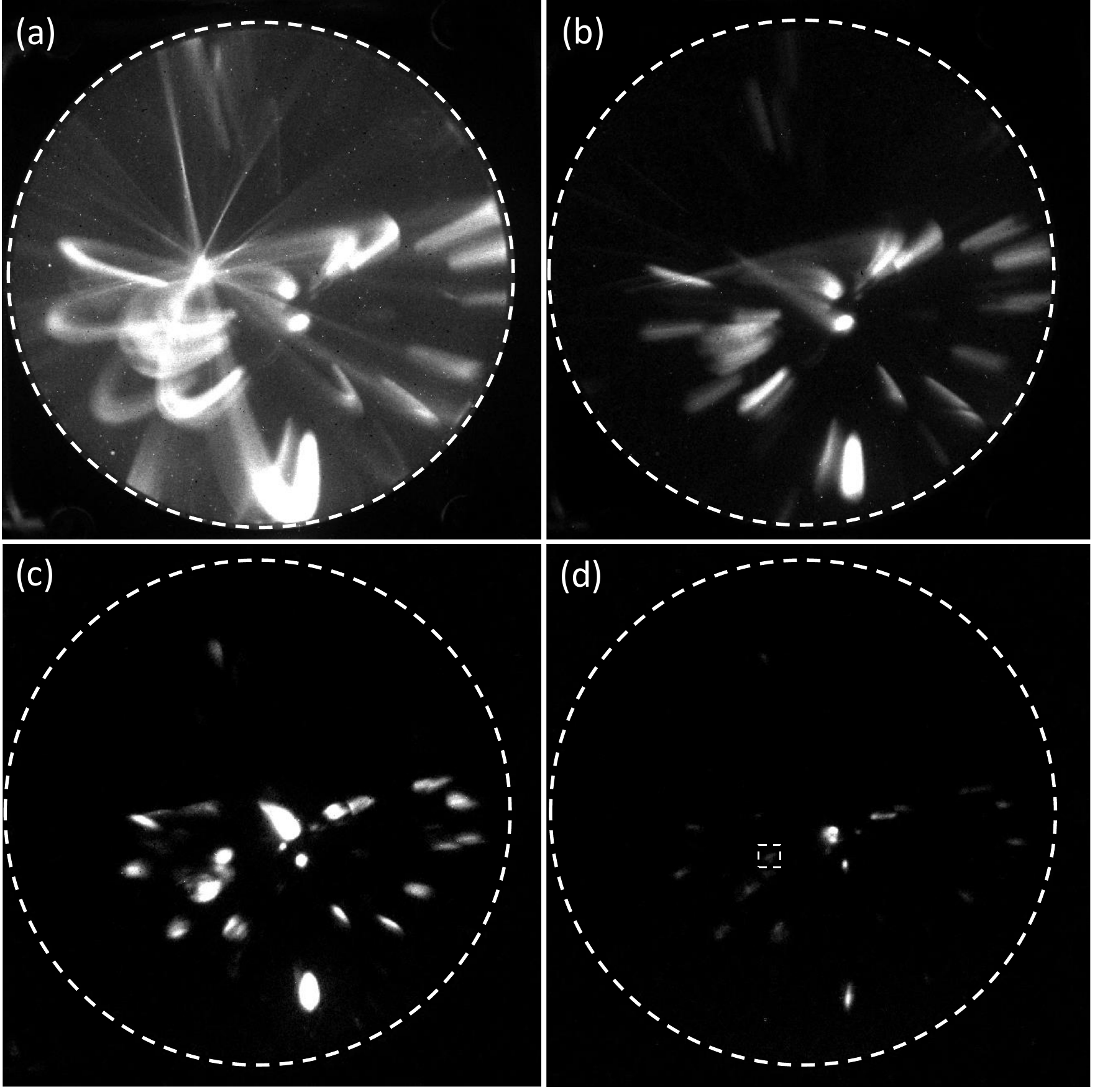}
	\caption{\label{figureApertures}Dark current images on the last YAG screen. The white dashed circle indicates the boundary of the YAG screen. The white dashed square in (d) indicates the emitter for the resolution calculation. (a) Without collimator. (b-d) With collimator. The aperture diameters are 8 mm, 1 mm, and 0.2 mm, respectively. (a-b) Accumulation of 20 frames. (c-d) Accumulation of 100 frames.}
\end{figure}

Typical dark current images on the last YAG screen are shown in Fig. \ref{figureApertures}. The imaging quality improves with smaller apertures, which validates the high-resolution dark current imaging method in rf structures by emitting phase and energy selection. The root-mean-square sizes of these emitters are determined by their actual sizes on the cathode as well as the system resolution. Taking a small emission area as marked by the white dashed square in Fig. \ref{figureApertures}(d), the axial and angular system resolution are calculated to be better than 147 $\mu$m and 107 $\mu$m, respectively. Emissions from unpredicted spots rather than the pre-defined pattern have been discovered. Most of these emitters were traced back to rf breakdown areas shown by the \emph{ex situ} surface examination. Despite a lower work function with respect to copper and gold, strong FE from the magnesium spots was not observed.

To date the Fowler-Nordheim equation is the most commonly used convention to quantitatively describe FE~\cite{FNformula1928,GadzukRMP1973,JuwenSLAC1997,Fursey2007}. Four determinants of the emission current are taken into account: the applied electric field strength, the emission area $A_{e}$, the material work function $\phi$, and the field enhancement factor $\beta$.  In previous studies of rf structures, $\beta$ is usually measured as an average value for a large surface~\cite{JuwenSLAC1997,HuaibiPRL2012,BaryshevAPL2014,JiahangPRL2015}. With the imaging system, $\beta$ can further measured for localized regions by quantifying their variation in luminous intensity with the rf field.

The brightness of the dark current image is proportional to the energy deposited on the YAG screen (luminance of the YAG screen has a linear response to the deposited energy). Along with other known parameters, the field enhancement factor $\beta$ of selected areas can be obtained by fitting to the F-N equation~\cite{JuwenSLAC1997}. During the measurement, $E_{c}$ was varied from 105 MV/m to 70 MV/m and the solenoid current was adjusted accordingly to maintain the same emitting positions on the last YAG screen. The biggest aperture (8 mm in diameter) was applied in this measurement to minimize the dependence of the capture ratio (defined as the current that can pass through the aperture divided by the total emission current from the cathode) on $E_{c}$. Other than FE electrons through the aperture, the brightness on the YAG screen also can be affected by the background luminance originating from X-rays generated by the energetic electrons, secondary electrons, light reflection along the beamline, etc. To quantify the background, dark current images were taken with a blank stainless steel plate at each field level. Then the background is subtracted from the image taken with the aperture present to ensure that the change of luminance value is only caused by the FE current through the aperture.

\begin{figure}[h!tbp]
\includegraphics[width=8cm]{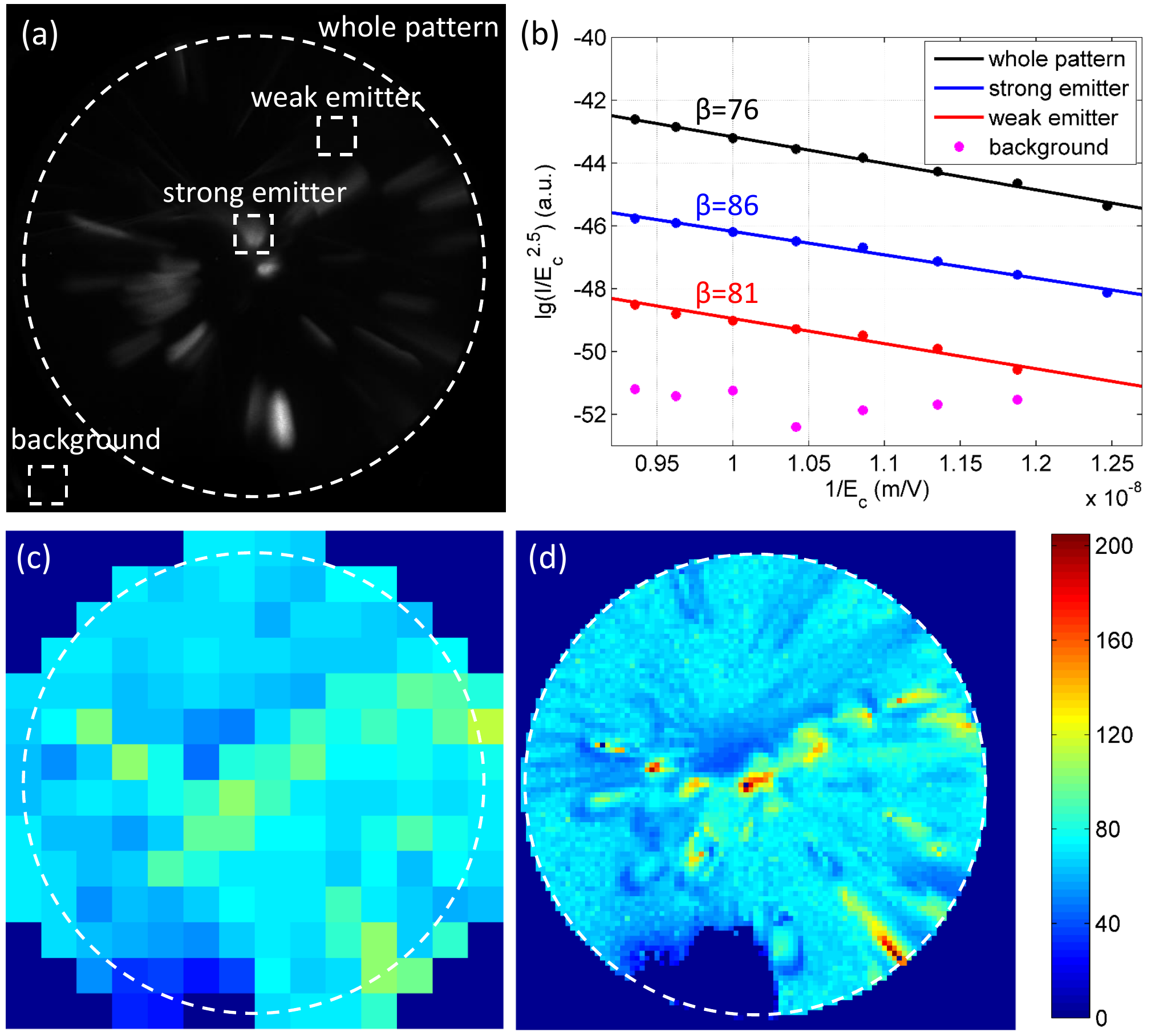}
\caption{\label{figureBeta}Field enhancement factor measurement by the dark current imaging system. The white dashed circle indicates the boundary of the YAG screen. (a) Dark current image with the biggest aperture. White dashed squares indicate selected regions for the measurement as shown in (b). (b) Fowler-Nordheim plot. Spots and lines are the measured data and linear fitting, respectively. (c-d) $\beta$ distribution map of the cathode with different selected region sizes. $\beta$ is set to zero for regions with non-linear dependence in the F-N plot. The size of select regions is 610 $\mu$m $\times$ 610 $\mu$m and 94 $\mu$m $\times$ 94 $\mu$m on the cathode for (c) and (d), respectively.}
\end{figure}

$\beta$ for the entire imaged area is 76 which falls into the typical range in previous studies~\cite{GrudievPRST2009,HuaibiPRL2012}. Region (610 $\mu$m $\times$ 610 $\mu$m on the cathode) which covers a strong emitter (bright spot) is selected for localized $\beta$ measurement. As a comparison, regions of the same size are also selected on the dark part inside and outside the YAG screen (illustrated in Fig. ~\ref{figureBeta}(a) as white dashed squares) for weak emitter and background, respectively. Results are shown in Fig. ~\ref{figureBeta}(b). The low value and non-linear dependence of data for the background confirms the brightness due to other sources has been subtracted correctly. However, $\beta$ of strong emitter and weak emitter is similar. In fact, while using such square size (610 $\mu$m $\times$ 610 $\mu$m on the cathode) the variation of $\beta$ over the whole surface is insignificant, as illustrated in Fig. ~\ref{figureBeta}(c).

\begin{figure}[h!tbp]
\includegraphics[width=8cm]{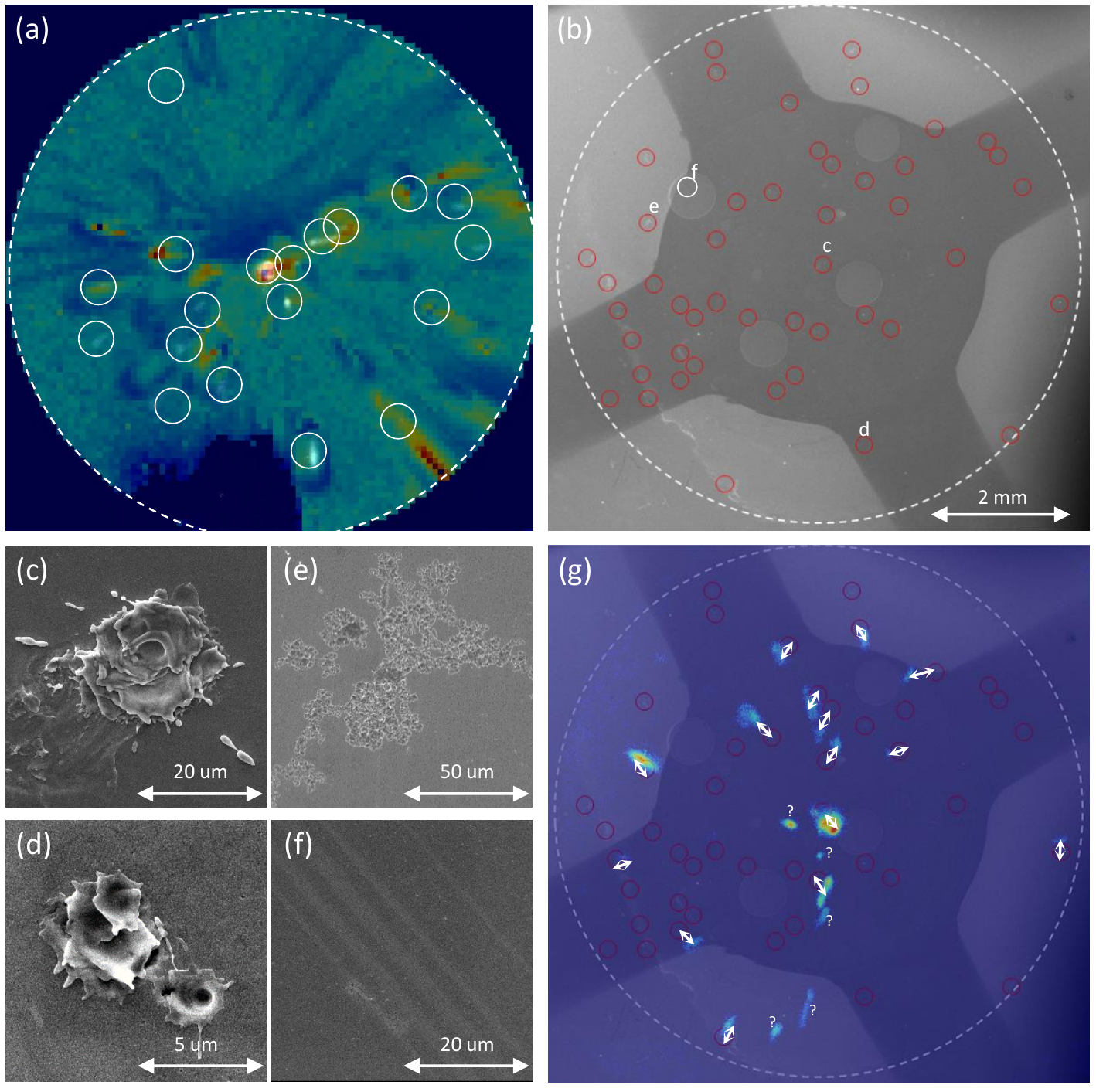}
\caption{\label{figureCompare}Overlap of high $\beta$ areas, strong dark current emitters and breakdown spots. (a) Overlap of high $\beta$ areas and strong dark current emitters. The white dashed circle indicates the
boundary of the YAG screen. The white solid circles indicate the areas which contain strong dark current emitters. (b) Overview of breakdown spots on the cathode. The red circles indicate the areas which contain breakdown spots. The white dash circle indicates the maximum visible range by the dark current imaging system. (c-e) Zoom-in view of circles marked in (b). (c) Breakdown spot on copper surface which overlaps with a strong dark current emitter. (d) Breakdown spot on copper surface which does not overlap with a strong dark current emitter. (e) Breakdown spot on gold which overlaps with a strong dark current emitter. (f) Smooth undamaged surface on magnesium. (g) Overlap of the strong dark current emitters and the breakdown spots. The dark current imaging is in false color for better display. The overlapped emitters and breakdown spots are marked by arrows. The emitters with unknown origin are marked by the question mark.}
\end{figure}

This is result of average effect of small effective emitter size (less than 200 $\mu$m in diameter deduced from Fig. ~\ref{figureApertures}(d)) and relatively large selected region for $\beta$ measurement. Much larger variation of $\beta$ over the cathode is obtained while using smaller regions (94 $\mu$m $\times$ 94 $\mu$m on the cathode) for measurement, as illustrated in Fig. ~\ref{figureBeta}(d). Most strong emission regions overlap with high $\beta$ areas, as illustrated in Fig. ~\ref{figureCompare}(a). A higher localized $\beta$ may be obtained with an improved imaging resolution.

After the imaging experiment, the cathode was examined by SEM and WLI as illustrated in Fig. ~\ref{figureCompare}(b-f). The major part of the surface remains intact after experiencing over hundreds of thousands of $\sim$100 MV/m rf pulses. The roughness of the smooth areas is 10-20 nm. Meanwhile, $\sim$40 breakdown spots have been observed within the areas as marked by the red circles in Fig. ~\ref{figureCompare}(b). Micro-structures such as melting craters and droplets are clearly signatures of the breakdown spots~\cite{JuwenSLAC1997}, which likely lead to high localized field enhancement. The microscopic valley deeper than 1 $\mu$m can be confirmed by the WLI. To study the relationship between the strong dark current emitters and the breakdown spots, the dark current image obtained with the smallest aperture has been resized and rotated based on the magnification and rotation from the cathode to the last YAG screen as simulated by the ASTRA code. The results show that $\sim$75\% of the strong emitters overlap with the breakdown spots, as illustrated in Fig. ~\ref{figureCompare}(g). The origin of the remaining $\sim$25\% strong emitters remains unknown. They may be attributed to microscopic surface features such as grain boundaries or defects that are not detected by the examination tools used. The results also reveal that half of the breakdown spots do not emit current at a level high enough to be detected by the imaging system.

The overlap of strong dark current emitters and breakdown spots with micro-structures supports the conventional understanding that FE may result from rf breakdowns in high gradient rf cavities~\cite{JuwenSLAC1997}. However the observation that no further breakdowns occurred at the strong dark current emitters imaged indicates that a steady FE alone may not be sufficient to trigger an rf breakdown.

In summary, a high-resolution \emph{in situ} dark current imaging setup has been developed based on an L-band photocathode gun. Separated emitters have been observed to dominate the field emission. The localized field enhancement factor has been measured. Post surface analysis by SEM and WLI reveals that $\sim 75\%$ strong dark current emitters overlap with the rf breakdown spots. This work greatly expands the understanding of field emission which in turn benefits research into electron sources, particle accelerators, and high gradient rf devices in general.

\begin{acknowledgments}
We would like to thank the Tsinghua University machine shop for preparing the new shaped cathodes, all staff in the AWA group for their work in the experiment, Dr. Klaus Fl\"{o}ttmann from DESY for his great help with the ASTRA code and other useful discussions, and Paul Schoessow of Euclid TechLabs for his valuable comments on the manuscript. The work by the AWA group is funded through the U.S. Department of Energy Office of Science under Contract No. DE-AC02-06CH11357. The work at Tsinghua University is supported by National Natural Science Foundation of China under Grant No. 11135004. The work by F. Wang is supported by the U.S. Department of Energy Early Career Research Program under Contract Code LAB 11-572. SEM measurements were conducted in the Electron Microscopy Center of the Center for Nanoscale Materials at Argonne National Laboratory. Use of the Center for Nanoscale Materials, an Office of Science user facility, was supported by the U. S. Department of Energy, Office of Science, Office of Basic Energy Sciences, under Contract No. DE-AC02-06CH11357.
\end{acknowledgments}

Gwanghui Ha is a visiting Ph.D. student at AWA from Pohang University of Science and Technology (POSTECH), South Korea.

\bibliography{imaging_ref}

\end{document}